\documentstyle[11pt]{article}

\begin{document}

{\bf Qualitative Analysis of Low-lying Resonances of the Dipositronium}

{\bf \ Emerging from Ps-Ps and Ps-Ps* Collisions}

{\bf \hspace{1.0in}}

BAO Cheng-guang$^{1}$ and SHI Ting-yun$^2$

\hspace{1.0in}

$^1$Department of Physics, Zhongshan University, Guangzhou, 510275

$^2$Wuhan Institute of Physics and Mathematics, The Chinese Academic of
Sciences, Wuhan, 430071


An analysis of the channel wave functions is made to clarify the types of
resonance emerging from the Ps-Ps and Ps-Ps* collisions . The ordering of
the energy levels of the states of the dipositronium is evaluated based on
the inherent nodal structures of wave functions and on existing theoretical
results. A few very probable low-lying narrower resonances, namely the 0$%
_1^{+}$(A$_2$), 1$_1^{-}$(E),$\cdot \cdot \cdot \cdot $, benefiting from the
centrifugal barrier have been proposed.

\hspace{1.0in}

Since the discovery of the positron the investigation of the systems
composed of positrons and electrons becomes an attractive topic. These
systems might exist in astrophysical processes and might play an important
role in astronomical evolution. Experimentally, the (e$^{+}$e$^{-}$) , (e$%
^{+}$e$^{-}$e$^{-}$), and (e$^{+}$e$^{+}$e$^{-}$) systems together with
their excited states have long been found in laboratories. However, the (e$%
^{+}$e$^{+}$e$^{-}$e$^{-}$) system, namely the dipositronium Ps$_2$, has not
yet been found. Theoretically, in recent years the existence of the ground
state of Ps$_2$ has been predicted by a number of authors$^{1-4}$, an
excited bound state with L$^\Pi =1^{-}$ (here L is the total orbital angular
momentum, $\Pi $ is the parity) has been predicted in the ref.[5], and two
more excited bound 0$^{+}$ states have also been predicted in ref.[1,6].
Obviously, in addition to these bound states, resonances of Ps$_2$ might
also exist, and they might emerge during Ps-Ps and Ps-Ps* collisions (here
Ps and Ps* denote the ground state and the excited state, respectively, of
the (e$^{+}$e$^{-}$) system). In this paper, a qualitative analysis is
performed to evaluate what types of low-lying resonances might be observed
during the above collisions.

Let us start from the Ps-Ps collision. Let the two positrons be called the
particles 1 and 2, while the two electrons be 3 and 4. Let us assume that
the spins of the positrons are coupled to $s_1$, and those of the electrons
are coupled to $s_2,$ and $\chi _{s_1s_2}$ is the spin state. Let the
spatial wave function of a Ps in its ground state be denoted as $\Phi $.
Then the antisymmetrized wave function $\Psi $ of the Ps-Ps channel can be
written as

$\Psi =\stackrel{\symbol{126}}{F}\chi _{s_1s_2}$\hspace{1.0in}(1)

where $\stackrel{\symbol{126}}{F}$ is a spatial wave function

$\stackrel{\symbol{126}}{F}=(1+(-1)^{s_1}p_{12})(1+(-1)^{s_2}p_{34})F$%
\hspace{1.0in}(2)

where $p_{ij}$ denotes an interchange of the indexes $i$ and $j$, and

$F=\Phi (\stackrel{\rightarrow }{r_{13}})\Phi (\stackrel{\rightarrow }{r_{24}%
})f_l(\stackrel{\rightarrow }{r}_{13,24})$\hspace{1.0in}(3)

where $\stackrel{\rightarrow }{r_{13}}=\stackrel{\rightarrow }{r_3}-%
\stackrel{\rightarrow }{r_1}$, $\stackrel{\rightarrow }{r}_{13,24}=\frac 12(%
\stackrel{\rightarrow }{r_2}+\stackrel{\rightarrow }{r_4})-\frac 12(%
\stackrel{\rightarrow }{r_1}+\stackrel{\rightarrow }{r_3}),$ etc.; and $%
\stackrel{\rightarrow }{r_i}$ is the position vector of the i-th particle
originating from the c.m.. $f_l$ is for the relative motion of the two Ps
with relative angular momentum $l$. Evidently, since the Ps are in their
ground states, $l$ is equal to the total orbital angular momentum L of the
channel. From (1) to (3) we have

$\stackrel{\symbol{126}}{F}=(1+(-1)^{s_1+s_2+{\rm L}})[\Phi (\stackrel{%
\rightarrow }{r_{13}})\Phi (\stackrel{\rightarrow }{r_{24}})f_L(\stackrel{%
\rightarrow }{r}_{13,24})+(-1)^{s_1}\Phi (\stackrel{\rightarrow }{r_{23}}%
)\Phi (\stackrel{\rightarrow }{r_{14}})f_L(\stackrel{\rightarrow }{r}%
_{23,14})]$\hspace{1.0in}(4)

It is obvious from (4) that the channel should have $s_1+s_2+$L$=even$.

When L=even, it is clear from (4) that

$p_{13}p_{24}\stackrel{\symbol{126}}{F}=\stackrel{\symbol{126}}{F}$%
\hspace{1.0in}(5)

Equation (5) implies that the spatial wave function $\stackrel{\symbol{126}}{%
F}$ of the channel should be invariant to $p_{13}p_{24}$ if L is even.

Furthermore, since the ground state of Ps has an even parity, the parity of
the channel $\Pi $ is therefore equal to (-1)$^l=(-1)^L$.

On the other hand, the eigenstates of the dipositronium Ps$_2$ can be
classified by the rotation group, space inversion group, and $D_{2d}$ point
group$^{1,7}$. Thus an eigenstate (bound state or resonance) can be labeled
as L$_i^\Pi (\mu )$, where $\mu $ denotes one of the representations A$_1$, A%
$_2$, B$_1$, B$_2$, and E of the $D_{2d}$ group; the subscript $i$ denotes
the $i-th$ state of a series of states with the same L, $\Pi ,$and $\mu $ .
The $i=1$ state is the lowest of the series. In some cases the subscript $i$
may be omitted if it is not necessary to appear. Incidentally, $\mu $ is
related to $s_1$ and $s_2$ as listed in Table 1$^7$. Furthermore, when $\mu
= $ A$_1$ and B$_1$, the spatial wave function $\stackrel{\symbol{126}}{F}$
of the L$_i^\Pi (\mu )$ states are invariant to $p_{13}p_{24}$ ; i.e., $%
\stackrel{\symbol{126}}{F}$ is an eigenstate of $p_{13}p_{24}$ with an
eigenvalue $\Lambda =1$. When $\mu =$ A$_2$ and B$_2$, $\stackrel{\symbol{126%
}}{F}$ is also an eigenstate of $p_{13}p_{24}$ but with an eigenvalue $%
\Lambda =-1$ .

Table 1

The relation between $\mu $ , $s_1$ and $s_2$, and $\Lambda $.

\begin{tabular}{|c|c|c|c|c|c|}
\hline
$\mu $ & A$_1$ & B$_2$ & B$_1$ & A$_2$ & E \\ \hline
$(s_1,s_2)$ & (0,0) & (0,0) & (1,1) & (1,1) & (0,1) or (1,0) \\ \hline
$\Lambda $ & 1 & -1 & 1 & -1 &  \\ \hline
\end{tabular}

Evidently, the good quantum numbers of a resonance should match those of the
channel wave functions. Thus, during the Ps-Ps collision, only the
resonances with

(i) $s_1+s_2+$L$=even$,

(ii) $\Lambda =1$ if L is even,

(iii) parity $\Pi =(-1)^{{\rm L}}$

can be induced . Consequently, three types of resonances , namely the L$%
_i^{+}($A$_1)$ states with an even L, the L$_i^{+}($B$_1)$ states with an
even L, and the L$_i^{-}($E$)$ states with an odd L, might emerge during
Ps-Ps collision. It is noted that the observation of a resonance depends on
its width. If the width is very broad, then the observation is difficult or
even impossible. Therefore, the above three types of resonances might not
all be observed.

The width of a resonance depends on how the wave function is distributed in
the channel region and in the interior. If the amplitude of the wave
function is small in the channel region (as to be compared with the
amplitude in the interior), then the width is narrow and the state is nearly
stable. If it is large in the channel region, then the width is broad. When
the width is broad, the resonance is difficult to be identified and
therefore might not be observed.

When the energy of a state is higher than the threshold of a channel and
when the quantum numbers of the state match those of the channel, there are
two factors to hinder the wave function in the interior from extending to
the channel region. One is the difference in structure between the channel
wave function and the wave function in the interior. When the difference is
large, the extension of the wave function is difficult, and therefore the
width is small. The other one is the centrifugal barrier and the Coulomb
barrier. These barriers play their role only if the bombarding energy is
low. Thus they are important to low-lying resonances in general. However,
since the Ps and Ps* are neutral in charge, the Coulomb barrier is
unimportant in our case, but the centrifugal barrier might be important. In
what follows we shall study the effect of this barrier to the low-lying
resonances.

Let us study the low-lying resonances with L$\leq 2$ induced by Ps-Ps
collision with the threshold energy -0.5 (atomic units are used in this
paper). It has been calculated$^1$ that the energies of the 0$_2^{+}$(A$_1$)
and 0$_1^{+}$(B$_1$) are -0.4995 and -0.4994, respectively . Both are a
little higher than the Ps-Ps threshold and therefore they might be observed
as resonances. However, since in this case there is no centrifugal barrier
to hinder the wave function, the widths of these two resonances might be
broad. Thus, even if the calculation in the ref. [1] is correct, these
resonances might not be observed.

The energy of the 1$_1^{-}$(E) state has never been calculated. Nonetheless,
it was known from all existing theoretical calculations that the energy of a
state would be high if the wave function contains many nodal surfaces; the
more the nodal surfaces, the higher the energy. Besides, it was found that
the first-states (the $i=1$ states) either do not contain any nodal surface,
or contain only the inherent nodal surfaces$^{8,9}$. Thus the inherent nodal
structures can be used to evaluate the ordering of the first-states. More
specifically, for the dipositronium, the ordering of the 0$_1^{+}$($\mu $)
levels have been found to be just ordered according to the number of
inherent nodal surfaces that they contain$^7$. This fact should also be true
for the group of other L$_1^\Pi $($\mu $) states with L$\neq $0. It was found%
$^7$ that the 1$_i^{-}$(E) states do not contain inherent nodal surfaces,
while all the other L=1 states contain inherent nodal surfaces$^7$. Thus the
1$_1^{-}$(E) state is the lowest L=1 state. It was calculated in the ref.[5]
that the energy of 1$_1^{-}$(B$_2$) is -0.3344. The energy of the 1$_1^{-}$%
(E) should be considerably lower than this value. If the energy of the 1$%
_1^{-}$(E) is lower than -0.5, then it is bound. If the energy is higher,
then it is a resonance. It is noted that the first excited L=0 state and the
lowest L=1 state of the positronium have exactly the same energy. This is in
general not true for other systems. However, the energies of the 0$_2^{+}$(A$%
_1$) and the 1$_1^{-}$(E) of the dipositronium might be close. If this is
true, the energy of the 1$_1^{-}$(E) would be close to -0.4995 . On the
other hand, the centrifugal barrier is equal to $\frac{l(l+1)}{2r^2}-0.5$,
where $l$ is the relative partial wave of the two Ps, and $r$ is the
relative distance of the two Ps. Once the channel radius has been evaluated,
the height of the barrier can be known. It is noted that the radius of a Ps
is equal to 2, the channel radius should be considerably larger than this
value. When the channel radius is given from 4 to 6, then the height of the $%
p-$wave barrier is from -0.4375 to -0.4722. So, if the energy of the 1$%
_1^{-} $(E) is considerably lower than the height of the barrier (e.g., the
energy is close to -0.4995), the collapse of the state would be effectively
hindered and accordingly the width would be narrower. In brief , there are
three possibilities. If the energy of the 1$_1^{-}$(E) is lower than -0.5,
the state is bound; if it is a little higher than -0.5, it is a narrow
resonance and thereby can be easily observed; if it is much higher than
-0.5, it might be a broad resonance and might not be observed.

The 2$_i^{+}$(A$_1$) and 2$_i^{+}$(B$_1$) states do not contain inherent
nodal surfaces, while the other 2$_i^\Pi $($\mu $) states contain$^7$. Thus
the 2$_1^{+}$(A$_1$) and 2$_1^{+}$(B$_1$) states are the two lowest L=2
states. It is very unlikely that they would be lower than the threshold at
-0.5, thus they are likely to be resonances. If their energies are
considerably lower than the height of the $d-$wave barrier (e.g., their
energies are lower than -0.46), they would have narrower widths and might be
observed. Otherwise, the widths might be broad and the observation might be
difficult.

In conclusion, the 1$_1^{-}$(E), 2$_1^{+}$(A$_1$) and 2$_1^{+}$(B$_1$) are
possible narrower low-lying resonances that might emerge in Ps-Ps collision.

Let us study the low-lying resonances with L$\leq 2$ induced by Ps-Ps*(2p)
collision with the threshold energy -0.3125, here the excited Ps is in the ($%
nl$)=(2p) state. In this case the spatial part of the channel wave function
reads

$\stackrel{\sim }{F}$=$\sum_l\{[\Phi (\stackrel{\rightarrow }{r_{13}})\Phi
_{2p}(\stackrel{\rightarrow }{r_{24}})+(-1)^{s_1+s_2+l}\Phi (\stackrel{%
\rightarrow }{r_{24}})\Phi _{2p}(\stackrel{\rightarrow }{r_{13}})]f_l(%
\stackrel{\rightarrow }{r_{13,24}})$

$+(-1)^{s_1}[\Phi (\stackrel{\rightarrow }{r_{23}})\Phi _{2p}(\stackrel{%
\rightarrow }{r_{14}})+(-1)^{s_1+s_2+l}\Phi (\stackrel{\rightarrow }{r_{14}}%
)\Phi _{2p}(\stackrel{\rightarrow }{r_{23}})]f_l(\stackrel{\rightarrow }{%
r_{23,14}})\}_L$\hspace{0.2in}(6)

where $\Phi _{2p}$ is the wave function of the Ps*(2p) . When ($%
s_1,s_2)=(0,0)$ or (1,1) , we have

$p_{13}p_{24}\stackrel{\sim }{F}=-\stackrel{\sim }{F}$\hspace{1.0in}(7)

Therefore, the Ps-Ps*(2p) channel can induce the L$_i^\Pi (\mu )$ resonances
with $\mu =A_2$ and $B_2$ (they have $\Lambda =-1$), and E$.$ Furthermore,
if L=0, we have $l=1$ and $\Pi =+1;$ therefore all the 0$_i^{-}(\mu )$
resonances can not be induced.

For the case of L=0, the $p-$wave of the Ps-Ps*(2p) collision can induce the
0$^{+}$(E), 0$^{+}$(B$_2$), and 0$^{+}$(A$_2$) resonances. However, it was
shown in the ref.[1] that the 0$_1^{+}$(E) and 0$_1^{+}$(B$_2$) states are
lower than the threshold, thus they can not be induced. Nonetheless, their
higher states (e.g., the 0$_2^{+}$(E) ) might be observed if they are higher
than and close to the threshold due to the $p-$wave barrier. The energy of
the 0$_1^{+}$(A$_2$) is -0.3121. Since the 0$_1^{+}$(A$_2$) is only a little
higher than the threshold, the $p$-wave barrier should act effectively. Thus
the width of this resonance should be narrow and therefore can be easily
observed. The search of this resonance is an interesting topic.

For the case of L=1, odd-parity resonances may be induced by s-wave while
even-parity resonances can only be induced by p-wave. Thus, even-parity
states in general might have a narrower width. Three kinds of even-parity
states, namely the 1$^{+}$(A$_2$), 1$^{+}$(B$_2$), and 1$^{+}$(E), can be
induced. It was found that the wave function of the 1$^{+}$(A$_2$) is
allowed by symmetry to be distributed surrounding a square with a pair of
the same kind of particles located at the two ends of a diagonal (this
geometric configuration is denoted as SQ), while the wave function of the 1$%
^{-}$(B$_2$) is not allowed$^7$. The SQ is the most favorable configuration
in favor of binding. Since the SQ is accessible to 1$^{+}$(A$_2$) but not
accessible to 1$^{-}$(B$_2$), the 1$_1^{+}$(A$_2$) would be lower than the 1$%
_1^{-}$(B$_2$). The latter has an energy -0.3344, thus the former would be
lower than the threshold and therefore might be bound. It was found$^7$ that
the 1$^{+}$(B$_2$) and 1$^{+}$($E$) contain more inherent nodal surfaces
than the 1$^{-}$(B$_2$) contains. Thus the 1$_1^{+}$(B$_2$) and 1$_1^{+}$($E$%
) would be higher than -0.3344. If their energies are higher than but close
to the threshold -0.3125, they would have narrower widths and therefore
might be observed. If they are much higher than the threshold, they would
have a broad width and therefore difficult to be observed.

For the L=2 resonances, the bombarding wave should at least have $l=1(2)$ if
the parity is even (odd). Since the energies of L=2 states have never been
evaluated, it is difficult to evaluate which resonances can be observed at
this moment. Nonetheless, the odd parity resonances might have a larger
possibility to be observed due to having a higher centrifugal barrier. Among
the odd states the 2$^{-}$(B$_2$) and 2$^{-}$(A$_2$) should be very high due
to having so many inherent nodal surfaces$^7$, while the 2$^{-}$(E) should
be lower because the SQ is accessible to this state. If the 2$_1^{-}$(E) is
lower than -0.3125, it is bound because the Ps-Ps channel is close to it; if
it is higher than and close to -0.3125, it would be a resonance with a
narrower width.

In conclusion, a number of L$_i^\Pi $($\mu $) resonances with $\mu =$A$_2$, B%
$_2$, and E might be induced in Ps-Ps*(2p) collision (among them the L=0
resonances have an even parity). Where the most probable narrower low-lying
resonance is the 0$_1^{+}$(A$_2$). The existence of the other suggested
narrower low-lying resonance, namely the 1$_1^{+}$(B$_2$), 1$_1^{+}$($E$) ,
and 2$_1^{-}$($E$) , is also possible

Let us study the low-lying resonances with L$\leq 2$ induced by Ps-Ps*(2s)
collision with also the threshold energy -0.3125. From an analysis similar
to the above, L$^\Pi $($\mu $) resonances with $\Pi =(-1)^{{\rm L}}$ and $%
\mu =$A$_1$, B$_1$, and E might be induced. However, the L=0 resonances do
not contain centrifugal barrier, thus they have broad widths. The 1$_1^{-}$(A%
$_1$) and 1$_1^{-}$(B$_1$) contain many inherent nodal surfaces just as the 0%
$_1^{+}$(A$_2$) state$^7$, but the energies of the former two should be
considerably higher than the latter (at -0.3121) due to the difference in L,
thus the former should be considerably higher than the threshold -0.3125.
Thus they can not take advantage from the $p-$wave barrier, and therefore
have a broad width. The 1$_1^{-}$(E) has been mentioned that it might appear
as a narrower resonance emerging from Ps-Ps collision. It is possible that
the 1$_2^{-}$(E) resonance might be observed during the Ps-Ps*(2s)
collision. Among the L=2 states, the 2$_1^{+}$(E) has the same inherent
nodal structure just as the 1$_1^{-}$(B$_2$)$^7$. The former should
considerably higher than the latter (at -0.3344) due to the difference in L,
but may not much higher than the threshold -0.3125. Thus the 2$_1^{+}$(E)
may take advantage from the $d-$ wave barrier and therefore has a narrower
width, and thereby can be observed.

In summary, an analysis based on symmetry has been performed in this paper.
Which types of resonance would emerge during the Ps-Ps and Ps-Ps* collisions
is clarified via the analysis of the channel wave functions. The ordering of
the energy levels of the states of the dipositronium is evaluated based on
the inherent nodal structures of wave functions and on existing theoretical
results. How the widths are affected by the centrifugal barrier has been
discussed. Emphasis is placed on the states that might have an energy higher
than and close to the Ps-Ps or Ps-Ps* threshold. A few very probable
low-lying narrower resonances benefiting from the centrifugal barrier have
been proposed. The existence of these suggested narrower resonances remain
to be checked.

Although the above qualitative analysis can only talk about the
possibilities but does not give definite quantitative results, the analysis
is still useful because it can help us to understand the physics underlying
the forthcoming experimental data and theoretical results. Specificallly, as
a first step, the experiment for the identification of the 0$_1^{+}$(A$_2$)
resonance via the Ps-Ps*(2p) collision is suggested. If this resonance can
not be found, this state would be the fifth bound state (together with the 0$%
_1^{+}$(A$_1$),0$_1^{+}$(E), 0$_1^{+}$(B$_2$), and 1$_1^{-}$(B$_2$) that
have been theoretically confirmed to be bound ).

Supported by the National Natural Science Foundation of China under Grant
No.19875084.

\hspace{1.0in}

{\bf REFERENCES}

1, D.B.Kinghorn and R.D. Poshusta, Phys. Rev. A 47 (1993) 3671

2, P.M.Kozlowski and L.Adamowicz, Phys. Rev. A48 (1993) 1903

3, A.M.Frolov and V.H.Smith, Jr., J. Phys. B29 (1996) L433; Phys. Rev. A55
(1997) 2662.

4, D.Bressanini, M.Mella, and G.Morosi, Phys. Rev. A55 (1997) 200

5, K.Varga, J.Usukura, and Y. Suzuki, Phys. Rev. Lett. 80 (1998) 1876

6, C.G.Bao and T.Y. Shi, Chin. Phys. Lett. 16 (1999) 267

7, C.G.Bao, Phys. Lett. A 243 (1998) 215

8, C.G. Bao, Few-Body Systems, 13 (1992) 57; Phys. Rev. A47 (1993) 1752.

9, C.G.Bao and Y.X. Liu, Phys. Rev. Lett. 82 (1999) 61

\end{document}